# Direct numerical simulation of compressible turbulence accelerated by graphics processing unit. Part 1: An open-source high accuracy accelerated computational fluid dynamic software.

Guanlin Dang (党冠麟)[a,c], Shiwei Liu (刘世伟)[b,c], Tongbiao Guo (郭同彪)[a], Junyi Duan (段俊亦)[a,c] and Xinliang Li (李新亮)[a,c,*]

[a]LHD, Institute of Mechanics, Chinese Academy of Sciences, Beijing 100190, China
[b]Academy of Mathematics and Systems Science, Chinese Academy of Sciences, Beijing 100190, China
[c]University of Chinese Academy of Sciences, Beijing 100049, China

ABSTRACT

This paper introduces open-source computational fluid dynamics software named open computational fluid dynamic code for scientific computation with graphics processing unit (GPU) system (OpenCFD-SCU), developed by the authors for direct numerical simulation (DNS) of compressible wall-bounded turbulence. This software is based on the finite difference method and is accelerated by the use of a GPU, which provides an acceleration by a factor of more than 200 compared with central processing unit (CPU) software based on the same algorithm and number of message passing interface (MPI) processes, and the running speed of OpenCFD-SCU with just 512 GPUs exceed that of CPU software with 130 000 CPUs. GPU-Stream technology is used to implement overlap of computing and communication, achieving 98.7% parallel weak scalability with 24 576 GPUs. The software includes a variety of high-precision finite difference schemes, and supports a hybrid finite difference scheme, enabling it to provide both robustness and high precision when simulating complex supersonic and hypersonic flows. When used with the wide range of supercomputers currently available, the software should able to improve the performance of large-scale simulations by up to two orders on the computational scale. Then, OpenCFD-SCU is applied to a validation and verification case of a Mach 2.9 compression ramp with mesh numbers up to 31.2 billion. More challenging cases using hybrid finite schemes are shown in Part 2(Dang, Li et al. 2022). The code is available and supported at http://developer.hpccube.com/codes/danggl/opencfd-scu.git.

## 1. Introduction

Compressible wall-bounded turbulent flows have received a considerable amount of research attention in recent years for their technological importance in aerospace applications such as supersonic or hypersonic vehicles and propulsion systems. In such research, experiments have historically played a major role, with numerical simulations only becoming common in the past two decades. Among all kinds of numerical simulation methods, direct numerical simulations (DNS) can provide turbulence data with the highest accuracy and resolution, with no turbulence model being required, and they have become a powerful tool for studying compressible turbulence. DNS have been reported to be capable of accurately capturing complex flow phenomena in supersonic or hypersonic internal and external flows, such as strong shock waves [1] and flow separation [2] and their interactions [3].

There are two significant challenges that have hindered the development of DNS of supersonic and hypersonic wall-bounded flows. The first concerns the numerical schemes adopted, especially for compressible turbulence with strong shock waves, where shock wave/boundary-layer interaction (SWBLI) occurs. Shock waves are ubiquitous in the flows around supersonic and hypersonic vehicles, with flow discontinuities occurring at the inlets of supersonic engines, on control surfaces, etc. A shock brings about a strong adverse pressure gradient that can lead to unsteady flow separation and unsteady motion of the shock, characterized by a wide range of frequencies. When a DNS is conducted, the numerical scheme needs to satisfy the conditions of numerical stability in the discontinuous region and high resolution and low dissipation in the whole flow domain. For flows at high Mach number (generally exceeding 6), a numerical dissipation scheme is usually introduced to preserve robustness where strong shock waves occur. Meanwhile, the dissipation in this scheme should be as low as as possible to maximize the accuracy of resolution. With increasing Mach number, the conflict between numerical stability and accurate resolution becomes more acute, which imposes stringent requirements on computational algorithms. The second challenge lies in the demand for huge computational resources. For example, to obtain a fully developed turbulent boundary layer flow, the computational domain must be extremely long, which makes accurate DNS extremely computationally demanding. This situation becomes more severe with increasing Reynolds number, when the grid size needs to be small enough to enable the description of turbulent vortices at all scales. In addition, the time step should be very small to capture the multiscale characteristics of turbulent fluctuations with time. Meanwhile, the total simulation time must be long enough to obtain the low-frequency unsteady motion of the shock, which has attracted much attention in recent years, with the mechanism by which such motion is generated still being the subject of much debate. Thus, DNS of supersonic and hypersonic flows requires huge computational resources, imposing further demands on ultra-large-scale parallel computing technology.

*Corresponding author
ORCID(s):

To improve both computational robustness and accuracy of resolution, hybrid finite difference schemes have been studied. In 2000, Adams [4] was the first to perform DNS of a supersonic turbulent boundary layer with a 18° turning angle of the compression ramp. He used a hybrid compact upwind–essentially nonoscillatory (upwind-ENO) scheme to discretize the convection term, enabling the calculation to deal with the discontinuity smoothly in the shock region while maintaining high precision in the relatively smooth region. Pirozzoli and Grasso [5] conducted a DNS to examine the interaction of an oblique shock wave with a flat plate turbulent boundary layer at Mach 2.25. They adopted a seventh-order weighted essentially nonoscillatory (WENO) scheme to discretize the convection terms. Although this WENO scheme succeeded in capturing the shock wave, the numerical dissipation that was introduced could have led to a reduction in computational accuracy. In 2007, Wu and Martín [6] studied the 24° compression ramp problem at Mach 2.9, improving the WENO scheme and constructing a seventh-order so-called WENO-SYMBO scheme [7] using the same grid base points as that of an eighth-order central scheme. In recent years, Ren et al. have proposed several finite difference schemes, including a minimized dispersion and controllable dissipation (MDCD) scheme [8] and a minimum dispersion and adaptive dissipation (MDAD) scheme [9] that have good dispersive and dissipative properties, which are important for the realization of high-fidelity simulations. They then combined these proposed schemes with the corresponding WENO schemes to provide shock-capturing capability.

Since the appearance of graphics processing units (GPUs), their floating-point computing power has more than doubled every two years, according to Huang's law [10]. By contrast, it is difficult for the number of transistors in the core of a central processing unit (CPU) to double every two years, as stated in Moore's law [11]. The computational performance of GPUs now far exceeds that of CPUs. From the TOP500 List of supercomputers [12] in May 2022 (see Table 1), it can be seen that eight of the top ten supercomputers have adopted GPUs as hardware accelerators. Using GPUs for program acceleration has become part of the mainstream of scientific and high-performance computing, especially for calculations that impose a high demand on computational resources. According to the GREEN500 List [13], only one supercomputer in the top 10 does not apply GPU acceleration.

Based on the above description, we have developed open-source computational fluid dynamics (CFD) software called OpenCFD-SCU with high precision and high performance, which includes hybrid finite difference schemes to improve the robustness of program computing and GPU acceleration that supports large-scale parallel computing. OpenCFD-SCU supports cross-platform deployment (for example, with either NVIDIA or AMD GPUs), and it has been optimized and tested to ensure its performance. It achieves 98.7% parallel weak scalability on 24 567 GPUs. Compared with the corresponding optimized CPU programs, the same number of Message Passing Interface (MPI) processes can be accelerated by a factor of more than 200, and the running speed of OpenCFD-SCU with just 512 GPUs exceed that of CPU software with 130 000 CPUs due to parallel efficiency decreasing with the increase of MPI processes.

The main purpose of this work is to describe the main features of OpenCFD-SCU, presenting a suite of validation and verification test cases, including supersonic and hypersonic compressible wall-bounded turbulence with shock waves. In addition, the optimization techniques can also provide a reference for other heterogeneous program development. The remainder of the paper is organized as follows. Section 2 gives an overview of the governing equations and numerical methods used in OpenCFD-SCU. In Section 3, the program framework and parallel implementation of the software are described. Section 4 introduces the hardware environment and the implementation and optimization techniques of OpenCFD-SCU using GPU acceleration. In Section 5, the accuracy of the software and its performance are assessed. In Section 6, a validation and verification case of a Mach 2.9 compression ramp with mesh number up to 31.2 billion are presented, and more challenging cases using hybrid finite schemes are shown in Part 2(Dang, Li et al. 2022). Finally, some conclusions are drawn in Section 7.

DNS of compressible turbulence accelerated by GPU

Table 1: Top 10 from the TOP500 list of supercomputers, as of May 2022 [12].

| Rank | Name | GPU used? | GPU model |
|---|---|---|---|
| 1 | Frontier | Yes | AMD MI250X |
| 2 | Fugaku | No | None |
| 3 | LUMI | Yes | AMD MI250X |
| 4 | Summit | Yes | NVIDIA GV100 |
| 5 | Sierra | Yes | NVIDIA GV100 |
| 6 | Sunway TaihuLight | No | None |
| 7 | Perlmutter | Yes | NVIDIA A100 |
| 8 | Selene | Yes | NVIDIA A100 |
| 9 | Tianhe-2A | No | None |
| 10 | Adastra | Yes | AMD MI250X |

## 2. Numerical algorithm

This section introduces the governing equations solved by OpenCFD-SCU and the algorithms included in the software. The program uses the finite difference method to solve the compressible three-dimensional Navier–Stokes equations in curvilinear coordinates. It can deal with a single-block grid in arbitrary curvilinear coordinates. The software integrates a variety of typical difference schemes and unique hybrid schemes. The hybrid schemes have good numerical stability, and they can simultaneously provide high accuracy and low numerical dissipation. Flux reconstruction using characteristic variables is also included as one of the options to enhance computational robustness, and it can be used when dealing with challenging problems.

### 2.1. Governing equations

The program solves the compressible three-dimensional nondimensionalized Navier–Stokes equations in curvilinear coordinates. The three-dimensional compressible Navier–Stokes equations in a curvilinear coordinate system can be written as

$$\frac{\partial \widetilde{U}}{\partial \tau} + \frac{\partial (\widetilde{F} - \widetilde{F}_v)}{\partial \xi} + \frac{\partial (\widetilde{G} - \widetilde{G}_v)}{\partial \eta} + \frac{\partial (\widetilde{H} - \widetilde{H}_v)}{\partial \varsigma} = 0, \quad (1)$$

where $\widetilde{F}$, $\widetilde{G}$, and $\widetilde{H}$ are the inviscid fluxes in the $\xi$, $\eta$, and $\varsigma$ directions, respectively, and $\widetilde{F}_v$, $\widetilde{G}_v$, and $\widetilde{H}_v$ are the corresponding viscous fluxes.

The transformation between the curvilinear coordinate system $(\xi, \eta, \varsigma, \tau)$ and the Cartesian coordinate system $(x, y, z, t)$ is given by

$$\begin{aligned} \xi &= \xi(x, y, z, t), \\ \eta &= \eta(x, y, z, t), \\ \varsigma &= \varsigma(x, y, z, t), \\ \tau &= t. \end{aligned} \quad (2)$$

Under the assumption that $\tau$ is a function only of $t$ and is independent of $x$, $y$, and $z$, the coordinate transformation satisfies the following relations:

$$\begin{aligned} \xi_x &= J(y_\eta z_\varsigma - z_\eta y_\varsigma), \quad \varsigma_x = J(y_\xi z_\eta - z_\xi y_\eta), \\ \xi_y &= J(z_\eta x_\varsigma - x_\eta z_\varsigma), \quad \varsigma_y = J(z_\xi x_\eta - x_\xi z_\eta), \\ \xi_z &= J(x_\eta y_\varsigma - y_\eta x_\varsigma), \quad \varsigma_z = J(x_\xi y_\eta - y_\xi x_\eta), \\ \eta_x &= J(y_\varsigma z_\xi - z_\varsigma y_\xi), \\ \eta_y &= J(z_\varsigma x_\xi - x_\varsigma z_\xi), \\ \eta_z &= J(x_\varsigma y_\xi - y_\varsigma x_\xi), \end{aligned} \quad (3)$$

where $J$ is the Jacobian of the coordinate transformation:

$$J = \left| \frac{\partial(\xi, \eta, \varsigma)}{\partial(x, y, z)} \right| = \left| \frac{\partial(x, y, z)}{\partial(\xi, \eta, \varsigma)} \right|^{-1}. \quad (4)$$

Therefore, the conservative variables $U$ in the Navier–Stokes equations are

$$\widetilde{U} = J^{-1} U = J^{-1} \begin{bmatrix} \rho \\ \rho u \\ \rho v \\ \rho w \\ E \end{bmatrix}. \quad (5)$$

The inviscid fluxes after the coordinate transformation are

$$\widetilde{F} = J^{-1} \begin{bmatrix} \rho \widetilde{u} \\ \rho u \widetilde{u} + \xi_x p \\ \rho v \widetilde{u} + \xi_y p \\ \rho w \widetilde{u} + \xi_z p \\ (E + p) \widetilde{u} \end{bmatrix}, \quad \widetilde{G} = J^{-1} \begin{bmatrix} \rho \widetilde{v} \\ \rho u \widetilde{v} + \eta_x p \\ \rho v \widetilde{v} + \eta_y p \\ \rho w \widetilde{v} + \eta_z p \\ (E + p) \widetilde{v} \end{bmatrix},$$

$$\widetilde{H} = J^{-1} \begin{bmatrix} \rho \widetilde{w} \\ \rho u \widetilde{w} + \varsigma_x p \\ \rho v \widetilde{w} + \varsigma_y p \\ \rho w \widetilde{w} + \varsigma_z p \\ (E + p) \widetilde{w} \end{bmatrix}, \quad (6)$$

where

$$\begin{aligned} \widetilde{u} &= \xi_t + \xi_x u + \xi_y v + \xi_z w, \\ \widetilde{v} &= \eta_t + \eta_x u + \eta_y v + \eta_z w, \\ \widetilde{w} &= \varsigma_t + \varsigma_x u + \varsigma_y v + \varsigma_z w \end{aligned} \quad (7)$$

and
$$E = \rho\left(\frac{p}{\gamma - 1} + \frac{u^2 + v^2 + w^2}{2}\right). \tag{8}$$

The viscous fluxes after the coordinate transformation are

$$\widetilde{F}_v = J^{-1}\begin{bmatrix} 0 \\ \xi_x\tau_{xx} + \xi_y\tau_{xy} + \xi_z\tau_{xz} \\ \xi_x\tau_{xy} + \xi_y\tau_{yy} + \xi_z\tau_{yz} \\ \xi_x\tau_{xz} + \xi_y\tau_{yz} + \xi_z\tau_{zz} \\ \xi_x\Theta_x + \xi_y\Theta_y + \xi_z\Theta_z \end{bmatrix}, \tag{9}$$

$$\widetilde{G}_v = J^{-1}\begin{bmatrix} 0 \\ \eta_x\tau_{xx} + \eta_y\tau_{xy} + \eta_z\tau_{xz} \\ \eta_x\tau_{xy} + \eta_y\tau_{yy} + \eta_z\tau_{yz} \\ \eta_x\tau_{xz} + \eta_y\tau_{yz} + \eta_z\tau_{zz} \\ \eta_x\Theta_x + \eta_y\Theta_y + \eta_z\Theta_z \end{bmatrix}, \tag{10}$$

$$\widetilde{H}_v = J^{-1}\begin{bmatrix} 0 \\ \varsigma_x\tau_{xx} + \varsigma_y\tau_{xy} + \varsigma_z\tau_{xz} \\ \varsigma_x\tau_{xy} + \varsigma_y\tau_{yy} + \varsigma_z\tau_{yz} \\ \varsigma_x\tau_{xz} + \varsigma_y\tau_{yz} + \varsigma_z\tau_{zz} \\ \varsigma_x\Theta_x + \varsigma_y\Theta_y + \varsigma_z\Theta_z \end{bmatrix}. \tag{11}$$

The stress work and heat transfer contributions are

$$\Theta_x = u\tau_{xx} + v\tau_{xy} + w\tau_{xz} + k\frac{\partial T}{\partial x},$$
$$\Theta_y = u\tau_{xy} + v\tau_{yy} + w\tau_{yz} + k\frac{\partial T}{\partial x}, \tag{12}$$
$$\Theta_z = u\tau_{xz} + v\tau_{yz} + w\tau_{zz} + k\frac{\partial T}{\partial x}.$$

The viscous stresses in a Newtonian fluid are

$$\tau_{ij} = \begin{cases} \mu\left(\dfrac{\partial u_j}{\partial x_j} + \dfrac{\partial u_j}{\partial x_i}\right), & i \neq j, \\ \mu\left(2\dfrac{\partial u_i}{\partial x_i} - \dfrac{2}{3}\operatorname{div} V\right), & i = j. \end{cases} \tag{13}$$

The heat conduction coefficient here is
$$k = \frac{C_p \mu}{\Pr \operatorname{Re}} = \frac{\mu}{(\gamma - 1)\operatorname{Ma}^2 \Pr \operatorname{Re}} \tag{14}$$

The nondimensionalized viscosity coefficient $\mu$ is given by Sutherland's formula
$$\mu = \frac{1 + C/T_{\text{ref}}}{T + C/T_{\text{ref}}}T^{3/2}, \tag{15}$$

where $C$ is the nondimensionalized speed of sound, which is given by
$$C^2 = \frac{T}{\operatorname{Ma}^2}. \tag{16}$$

In the program, the dimensionless parameters are expressed as

$$\begin{aligned} x &= x_{\text{local}}/L_{\text{ref}}, & u &= u_{\text{local}}/U_{\text{ref}}, \\ t &= t_{\text{local}}U_{\text{ref}}/L_{\text{ref}}, & \rho &= \rho_{\text{local}}/\rho_{\text{ref}}, \\ T &= T_{\text{local}}/T_{\text{ref}}, & p &= p_{\text{local}}/(\rho_{\text{ref}}U_{\text{ref}}^2). \end{aligned} \tag{17}$$

The solver only needs to be given the parameters $T_{\text{ref}}$, $\gamma$, Re, Ma, and Pr to carry out the simulation. The $\gamma$ of an ideal gas is generally 1.4, and Pr is generally 0.7, while Re and Ma can be calculated from other reference parameters:

$$\operatorname{Re} = \frac{\rho_{\text{ref}}U_{\text{ref}}L_{\text{ref}}}{\mu_{\text{ref}}}, \quad \operatorname{Ma} = \frac{U_{\text{ref}}}{C_{\text{ref}}}, \quad \text{with } C_{\text{ref}}^2 = \gamma R T_{\text{ref}}. \tag{18}$$

### 2.2. Numerical discretization for inviscid terms

The flow flux vector is first split to allow the inviscid term to be dealt with, and the original inviscid flux is decomposed into positive and negative fluxes. Then, the corresponding difference discretizations (usually upwind schemes) are adopted for these fluxes. Compared with direct discretization of the convection term without flux vector splitting (usually using a central scheme or spectral method), an upwind difference scheme after flux vector splitting has better stability. It can effectively suppress confusion error [14]. In addition, for flows with shocks or discontinuities, flow flux vector splitting combined with a shock capture scheme can effectively supress nonphysical oscillations of the numerical solution, thereby achieving a good simulation result.

#### 2.2.1. Steger–Warming split

OpenCFD-SCU uses the Steger-Warming [15] flux vector splitting method, which is widely used in hypersonic flow.

The three-dimensional flux vector is expressed as a sum of the flux vectors in three directions:
$$f = \alpha_1 f_1 + \alpha_2 f_2 + \alpha_3 f_3. \tag{19}$$

In the calculation of the fluxes in the $x$ direction ($\alpha_1 = \xi_x/J, \alpha_2 = \xi_y/J, \alpha_3 = \xi_z/J$), the fluxes in the $y$ direction ($\alpha_1 = \varsigma_x/J, \alpha_2 = \varsigma_y/J, \alpha_3 = \varsigma_z/J$), and the fluxes in the $z$ direction ($\alpha_1 = \eta_x/J, \alpha_2 = \eta_y/J, \alpha_3 = \eta_z/J$), the Jacobian matrix $A = D(f)/D(U)$ is used, whose eigenvalues are

$$\lambda_1 = \lambda_2 = \lambda_3 = \widetilde{V}, \quad \lambda_4 = \widetilde{V} - c\sigma, \quad \lambda_5 = \widetilde{V} + c\sigma, \tag{20}$$

where
$$\widetilde{V} = \alpha_1 u + \alpha_2 v + \alpha_3 w, \tag{21}$$
$$\sigma = \sqrt{\alpha_1^2 + \alpha_2^2 + \alpha_3^2} \tag{22}$$

The three-dimensional flow vector $\boldsymbol{f}$ is expressed as

$$\boldsymbol{f} = \frac{\rho}{2\gamma} \begin{bmatrix} 2(\gamma-1)\widetilde{\lambda}_1 + \widetilde{\lambda}_4 + \widetilde{\lambda}_5 \\ 2(\gamma-1)\widetilde{\lambda}_1 u + \widetilde{\lambda}_4(u - c\widetilde{k}_1) + \widetilde{\lambda}_5(u + c\widetilde{k}_1) \\ 2(\gamma-1)\widetilde{\lambda}_1 v + \widetilde{\lambda}_4(v - c\widetilde{k}_1) + \widetilde{\lambda}_5(v + c\widetilde{k}_1) \\ 2(\gamma-1)\widetilde{\lambda}_1 w + \widetilde{\lambda}_4(w - c\widetilde{k}_1) + \widetilde{\lambda}_5(w + c\widetilde{k}_1) \\ (\gamma-1)\widetilde{\lambda}_1 V + W + \widetilde{\lambda}_4 v_{c1} + \widetilde{\lambda}_5 v_{c2} \end{bmatrix}, \tag{23}$$

with

$$v_{c1} = \frac{\widetilde{\lambda}_4}{2}\left[(u - c\widetilde{k}_1)^2 + (v - c\widetilde{k}_1)^2 + (w - c\widetilde{k}_1)^2\right],$$

$$v_{c2} = \frac{\widetilde{\lambda}_5}{2}\left[(u + c\widetilde{k}_1)^2 + (v + c\widetilde{k}_1)^2 + (w + c\widetilde{k}_1)^2\right]. \tag{24}$$

Here

$$V = \tfrac{1}{2}(u^2 + v^2 + w^2), \tag{25}$$

$$W = \frac{(3-\gamma)(\widetilde{\lambda}_4 + \widetilde{\lambda}_5)c^2}{2(\gamma-1)}, \tag{26}$$

$$\widetilde{k}_1 = \frac{\alpha_1}{\sigma_1}, \quad \widetilde{k}_2 = \frac{\alpha_2}{\sigma_2}, \quad \widetilde{k}_3 = \frac{\alpha_3}{\sigma_3}. \tag{27}$$

The eigenvalues are decomposed into positive and negative eigenvalues:

$$\lambda_k = \lambda_k^+ + \lambda_k^- \quad (k = 1, \ldots, 5), \tag{28}$$

$$\lambda_k^+ = \frac{\lambda_k + |\lambda_k|}{2}, \quad \lambda_k^- = \frac{\lambda_k - |\lambda_k|}{2}. \tag{29}$$

Insertion of (23) gives the positive and negative fluxes as

$$\boldsymbol{f} = \boldsymbol{f}^+ + \boldsymbol{f}^-. \tag{30}$$

$$\begin{array}{c} \cdots\cdots\;j-4\quad j-3\;\;j-2\;\;j-1\;\;j\;\;\overset{\bullet}{j+1}\;\;j+2\;\;j+3\;\cdots\cdots \\ \left.\dfrac{\partial f(U)}{\partial x}\right|_{j+1/2} \end{array}$$

Figure 1: Difference scheme stencil point.

### 2.2.2. Flux reconstruction using characteristic variables

When calculating the numerical flux, the flux is first converted to the characteristic space. The characteristic flux can then be reconstructed to enhance the robustness of the program and reduce nonphysical oscillations, although at the cost of a greater number of calculations. This approach can be chosen when dealing with complex cases.

Fig. 1 shows the stencil point required for the differencing. The derivative of the inviscid flux $\boldsymbol{f}$ is expressed at the stencil point as follows:

$$\left.\frac{\partial \boldsymbol{f}(\mathbf{U})}{\partial x}\right|_{j+1/2} = \mathbf{A}\left.\frac{\partial \mathbf{U}}{\partial x}\right|_{j+1/2} = \mathbf{S}^{-1}\mathbf{\Lambda}\mathbf{S}\left.\frac{\partial \mathbf{U}}{\partial x}\right|_{j+1/2}$$

$$= \mathbf{S}^{-1}\left.\frac{\partial(\mathbf{\Lambda}\mathbf{S}\mathbf{U})}{\partial x}\right|_{j+1/2}. \tag{31}$$

Here, $\mathbf{U}_{j+1/2}$ can be obtained by simple averaging:

$$\mathbf{U}_{j+1/2} = \frac{\mathbf{U}_j + \mathbf{U}_{j+1}}{2}. \tag{32}$$

The eigenvalues are the same as in (20), which here takes the form

$$\mathbf{\Lambda} = \sigma\,\mathrm{diag}(u_n, u_n, u_n, u_n - c, u_n + c). \tag{33}$$

There are many different ways to express the eigenmatrix, and the form adopted in this paper is

$$\mathbf{S}^{-1} = \begin{bmatrix} 1 & 0 & 0 & 1 & 1 \\ u & l_1 & m_1 & u - cn_1 & u + cn_1 \\ v & l_2 & m_2 & v - cn_2 & v + cn_2 \\ w & l_3 & m_3 & w - cn_3 & w + cn_3 \\ V & u_l & u_m & H - cu_n & H + cu_n \end{bmatrix}, \tag{34}$$

$$\mathbf{S} = \begin{bmatrix} 1 - \kappa V & \kappa u & \kappa v & \kappa w & -\kappa \\ -u_l & l_1 & l_2 & l_3 & 0 \\ -u_m & m_1 & m_2 & m_3 & 0 \\ \kappa_1 V + \chi u_n & -\chi n_1 - \kappa_1 u & -\chi n_2 - \kappa_1 v & -\chi n_3 - \kappa_1 w & \kappa_1 \\ \kappa_1 V - \chi u_n & \chi n_1 - \kappa_1 u & \chi n_2 - \kappa_1 v & \chi n_3 - \kappa_1 w & \kappa_1 \end{bmatrix}. \tag{35}$$

Here,

$$H = \frac{E + p}{\rho} = V + \frac{c^2}{\gamma - 1}, \tag{36}$$

$$\kappa = \frac{\gamma - 1}{c^2}, \quad \kappa_1 = \frac{\kappa}{2}, \quad \chi = \frac{1}{2c}, \tag{37}$$

$$\begin{aligned} u_n &= un_1 + vn_2 + wn_3, \\ u_l &= ul_1 + vl_2 + wl_3, \\ u_m &= um_1 + vm_2 + wm_3 \end{aligned} \tag{38}$$

$V$ is given by (25). $\boldsymbol{l} = (l_1, l_2, l_3)$, $\boldsymbol{m} = (m_1, m_2, m_3)$, and $\boldsymbol{n} = (n_1, n_2, n_3)$ are three pairwise-orthogonal unit vectors. There are many methods to select these vectors, and this program uses the following:

$$(l_1, l_2, l_3) = \begin{cases} (-n_2, n_1, 0)/\sqrt{n_1^2 + n_2^2} & \text{if } |n_3| \leq |n_2|, \\ (-n_3, 0, n_1)/\sqrt{n_1^2 + n_3^2} & \text{otherwise,} \end{cases} \quad (39)$$

$$\boldsymbol{m} = \boldsymbol{n} \times \boldsymbol{l} = (n_2 l_3 - n_3 l_2, n_3 l_1 - n_1 l_3, n_1 l_2 - n_2 l_1). \quad (40)$$

### 2.2.3. Difference scheme

The difference discretization of inviscid terms is generally an upwind difference scheme, with corresponding upwind differences being used for the split positive and negative flux vectors. A variety of upwind difference schemes are integrated into OpenCFD-SCU, as listed in Table 2. In addition to the typical difference schemes in Table 2, the program also integrates a hybrid scheme to calculate inviscid terms, which is an excellent way to solve complex problems. In the 1970s, Harten first put forward the concept of a hybrid scheme [16]. Later, Adams and Shariff [17] used a hybrid scheme to simulate shock–boundary layer interference in the case of a supersonic ramp angle for the first time. In the OpenCFD-SCU program, a modified Jameson shock sensor [18] is used to judge the flow field, and three appropriate difference schemes are selected for different parts.

Table 2: Difference schemes integrated into OpenCFD-SCU for inviscid terms.

| Scheme No. | Upwind difference scheme |
|---|---|
| 1 | Second-order NND [19] |
| 2 | Fifth-order WENO |
| 3 | Sixth-order OMP6 [20] |
| 4 | Seventh-order WENO [21] |
| 5 | Seventh-order WENO-SYMBO [7] |
| 6 | Seventh-order upwind |

Of the three schemes comprising the hybrid scheme, the first is a seventh-order upwind scheme:

$$f_{j+1/2}^{UD7L} = (-3.0 f_{j-3} + 25.0 f_{j-2} - 101.0 f_{j-1} + 319.0 f_j \\ + 214.0 f_{j+1} - 38.0 f_{j+2} + 4.0 f_{j+3})/420.0. \quad (41)$$

Compared with the general shock capture scheme, the seventh-order upwind scheme has lower dissipation and high accuracy, but it has a poor ability to deal with discontinuities. In the hybrid scheme, the seventh-order upwind scheme calculates most of the flow fields except discontinuities.

The second scheme is seventh-order WENO [21]. The WENO scheme is a high-precision shock capture scheme proposed based on high-precision ENO and using the idea of a weighted average. It is widely used in the simulation of physical problems. The seventh-order WENO scheme has acceptable dissipative characteristics and high accuracy, and because of its good robustness, it is often used to simulate supersonic turbulence. However, if we want to obtain good simulation data, the dissipation of seventh-order WENO is still too high. In the hybrid scheme, seventh-order WENO can be used to calculate some weak discontinuities in the flow field.

The third scheme is fifth-order WENO. Compared with seventh-order WENO, fifth-order WENO has more substantial dissipation, which means better robustness. It can calculate strong discontinuities and shock waves as part of the hybrid scheme.

The modified Jameson sensor is given by

$$\begin{aligned} \phi_i &= \frac{|-p_{i-1} + 2p_i - p_{i+1}|}{p_{i-1} + 2p_i + p_{i+1}}, \\ \phi_j &= \frac{|-p_{j-1} + 2p_j - p_{j+1}|}{p_{j-1} + 2p_j + p_{j+1}}, \quad (42) \\ \phi_k &= \frac{|-p_{k-1} + 2p_k - p_{k+1}|}{p_{k-1} + 2p_k + p_{k+1}}, \\ \theta &= \phi_i + \phi_j + \phi_k. \quad (43) \end{aligned}$$

The thresholds are then set as $\theta_1$ and $\theta_2$. When $\theta \leq \theta_1$, the seventh-order upwind scheme is used; when $\theta_1 < \theta \leq \theta_2$, seventh-order WENO is used, and when $\theta_2 \leq \theta$, fifth-order WENO is used. The higher the values of $\theta_1$ and $\theta_2$, the higher is the dissipation of the hybrid scheme. They are usually set to 0.02 and 0.1, respectively.

This hybrid scheme has good numerical stability for complex cases. When used in combination with flux reconstruction using characteristic variables, the whole simulation process can be guaranteed not to diverge, provided the time step is appropriate. Moreover, in the simulation process, more than 90% of the positions use the low-dissipation seventh-order upwind scheme; only in a few locations where there is a shock wave or flow separation will the seventh- or fifth-order WENO be used. Because the accuracy of the three hybridized schemes is high (more than fifth-order), only their dissipation characteristics are different, and the pseudo-waves that may be generated after scheme switching are relatively slight. Even if these pseudo-waves are generated, they will be submerged by a local shock wave, turbulence, or flow separation, thus ensuring the accuracy of the calculation.

### 2.2.4. Boundary scheme

For an aperiodic boundary, order reduction is needed near the boundary. This program uses the fifth-order WENO scheme to calculate the numerical flux near the boundary. When a subtemplate of the WENO scheme uses grid points outside the boundary, the weight of the subtemplate is forced to 0 (shielding the out-of-boundary template).

### 2.3. Numerical discretization for viscous terms

A central difference scheme is generally adopted for the difference discretization of viscous terms, and two such schemes, of sixth and eighth order, respectively, are integrated into the program. For an aperiodic boundary, order reduction is adopted in the boundary calculation. A second-order unilateral difference scheme is used at the boundary point, and second-, fourth-, and sixth-order central difference schemes are used at the second, third, and fourth boundary points, respectively.

### 2.4. Time marching method

The three-step and third-order Runge–Kutta method is used to march the time.

For the equation

$$\frac{\partial U}{\partial t} = L(U), \tag{44}$$

the time marching process is as follows:

$$\begin{aligned} U^{(1)} &= U^n + \Delta t L(U^n), \\ U^{(2)} &= \tfrac{3}{4}U^n + \tfrac{1}{4}[U^{(1)} + \Delta t L(U^{(1)})], \\ U^{n+1} &= \tfrac{1}{3}U^n + \tfrac{2}{3}[U^{(2)} + \Delta t L(U^{(2)})]. \end{aligned} \tag{45}$$

## 3. Software

This section introduces the program framework and the parallel implementation of the software. The OpenCFD-SCU program adopts the heterogeneous programming model of C+CUDA. This program performs double-precision calculations, with the GPU carrying out almost all double-precision floating-point computing tasks and the CPU being responsible only for program logic control and data communication. The calculation stream of the program has been designed such that four streams are opened up through CUDA-Stream technology, and there is overlap of computing communications, which ensures excellent expansibility of the program in parallel implementations.

### 3.1. Program framework

Fig. 2 shows the program framework of OpenCFD-SCU. The program is initialized by reading external control parameters, and then the data memory is allocated to the CPU and GPU, respectively. The external data are first read into each CPU process through MPI parallel input and output (IO) and then passed to the GPU by the CPU. After this, the calculation begins. The operations in red boxes in the figure require data communication between processes, and the operations in blue boxes require data exchange between CPU and GPU. The program puts the updating of the interior and outer zones of the fluxes of the inviscid term and the viscous term on four streams. When updating the fluxes, the flux in the $x$ direction is updated first, and then the fluxes in the $y$ and $z$ directions are updated. After the flux update, time is marched, the original variable is updated, and the boundary condition is called. When the calculated time step reaches the maximum time step set, the program terminates the calculation, and the data are saved and displayed as previously determined by the user.

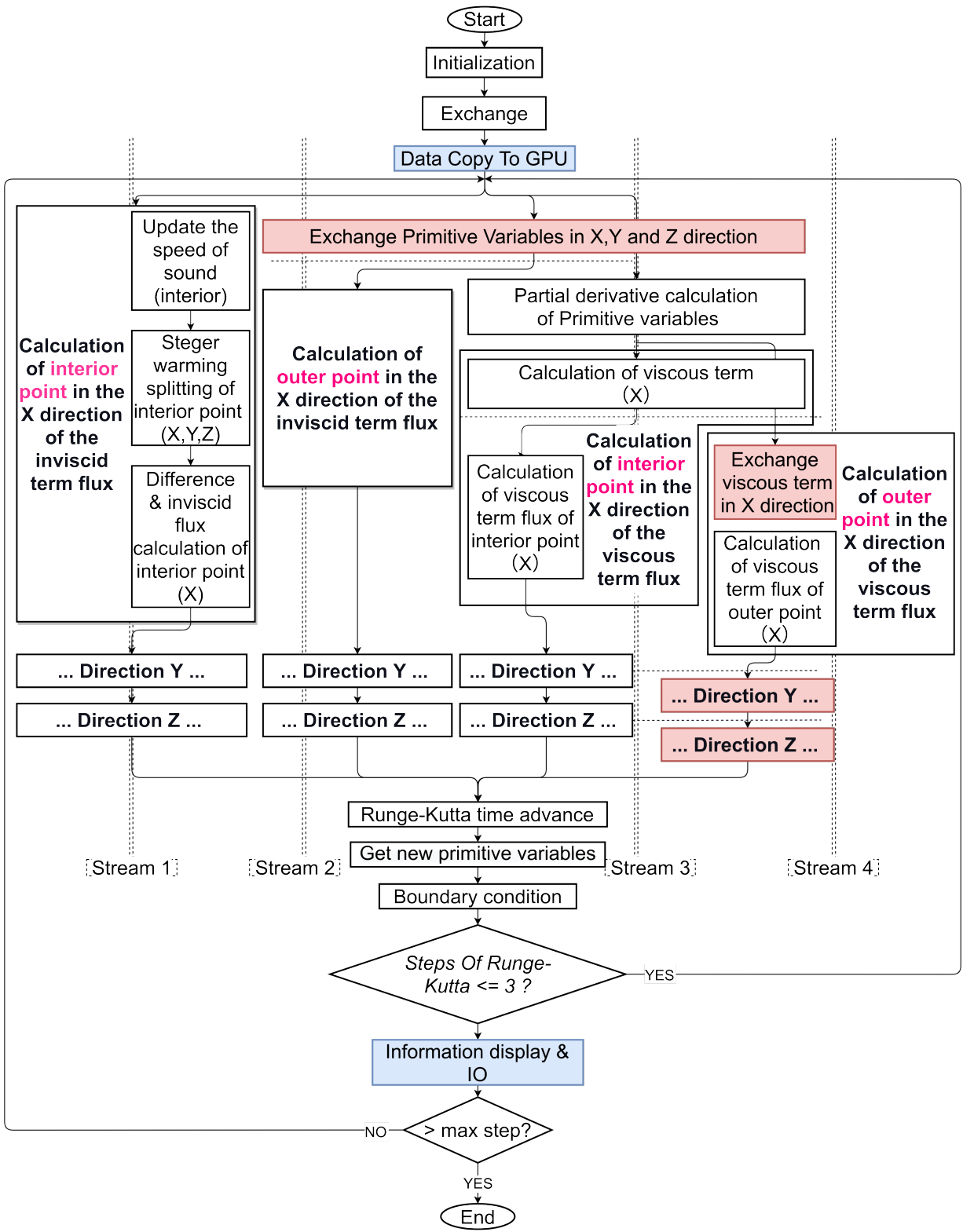

Figure 2: Program framework of OpenCFD-SCU.

## 3.2. MPI communication

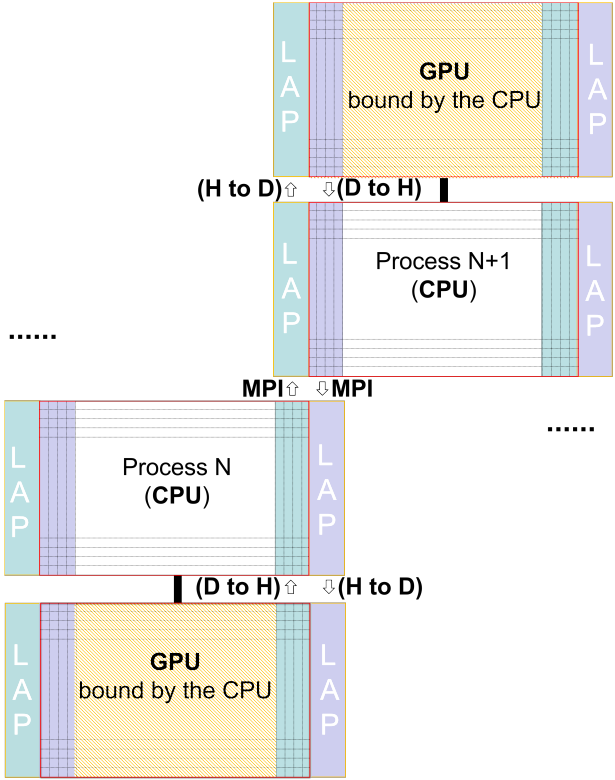

Figure 3: MPI interprocess data exchange mode of OpenCFD-SCU.

This program supports large-scale parallel computing. Its parallel programming model is Message Passing Interface (MPI), the most popular parallel programming message passing specification, which is supported by most supercomputing platforms.

At present, much of the available supercomputing hardware does not support GPU direct connection technology, which was first developed by NVIDIA to exchange data between GPUs and requires specific hardware support, and so this program does not use this technology. When the program carries out interprocess data communication, it needs to transfer data from GPU to CPU, and then the CPU communicates through MPI. The communication process is shown in Fig. 3. When communicating, the GPU does not need to transfer all the data to the CPU, just the data needed by the adjacent process stencil. The data that must be exchanged between processes constitute only a few thin layers outside the data block.

This procedure uses a three-dimensional MPI partition; the data division method is shown in Fig. 4. Each block represents data on each process, and the *MPI_Comm_split()* function generates the index information in three directions. The orange framed zones in the figure represent the overlapping zones between processes. Each process transfers several layers of data from its boundary to the lap zone of the adjacent process, and then obtains the data from the adjacent process and stores it in its lap zone. The outer zone of each data block is shown in pink. In the calculation, only the outer zone needs to obtain the adjacent process data through MPI communication, with the inner zone only needing to obtain the data locally. Therefore, the calculation of the inner zone can overlap with the calculation of the outer zone and the MPI communication.

Fig. 5 illustrates the data partition of a single process more clearly. For convenience, the memory requested by each process is a box of $(nx + 2\,\text{LAP}) \times (ny + 2\,\text{LAP}) \times (nz + 2\,\text{LAP})$, where LAP is the number of points in the lap zone width, but the actual useful memory, as shown in Fig. 5(a), is not a complete box. In this program, the highest supported difference scheme is of seventh order. Therefore, LAP is generally set equal to 4, and the thickness of the outer zone is set to double the width of the lap zone, which is for the merging and alignment of GPU data access. Fig. 5(b) is a middle section of Fig. 5(a), providing a clearer view of the lap zone, outer zone, and interior zone. Fig. 5(c) is a split chart of the data structure, from which it can be seen that the size of the outer zone is not the same in the three directions: the size in the $x$ direction is $\text{LAP} \times (ny - 4\,\text{LAP}) \times (nz - 4\,\text{LAP})$, the size in the $y$ direction is $nx \times \text{LAP} \times (nz - 4\,\text{LAP})$, and the size in the $z$ direction is $nx \times ny \times \text{LAP}$. These can be combined to wrap the interior zone points.

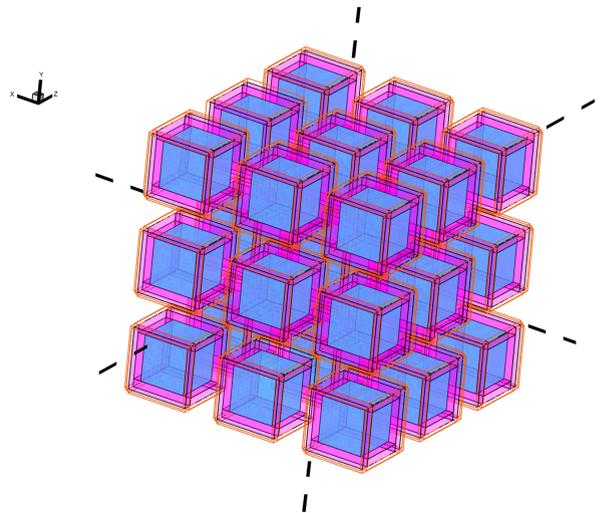

Figure 4: Data partition of 3D MPI process.

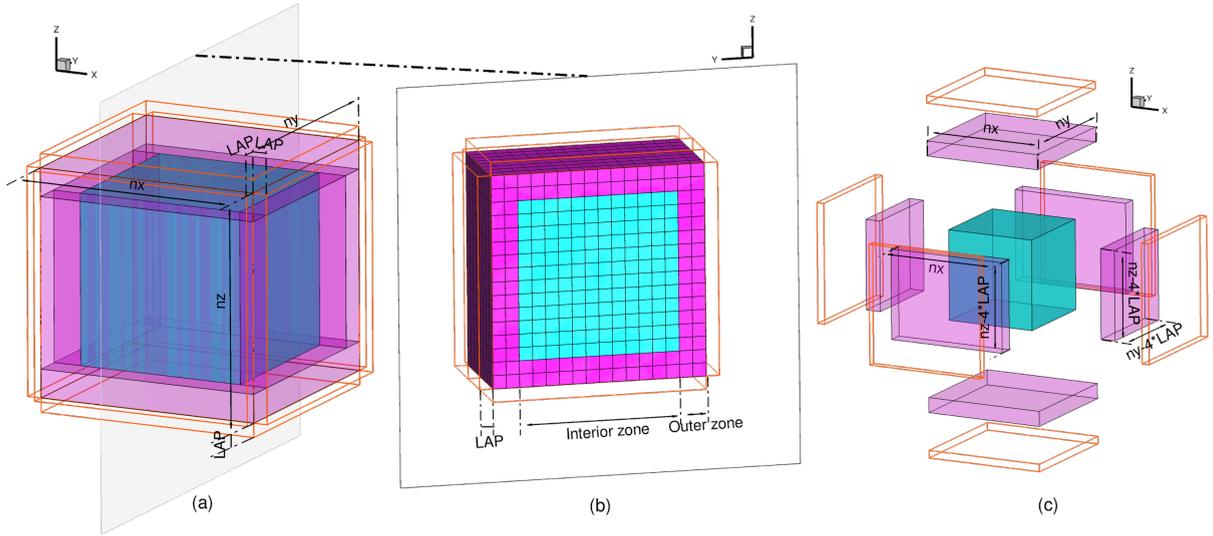

Figure 5: A data structure within an MPI process.

## 3.3. Stream design

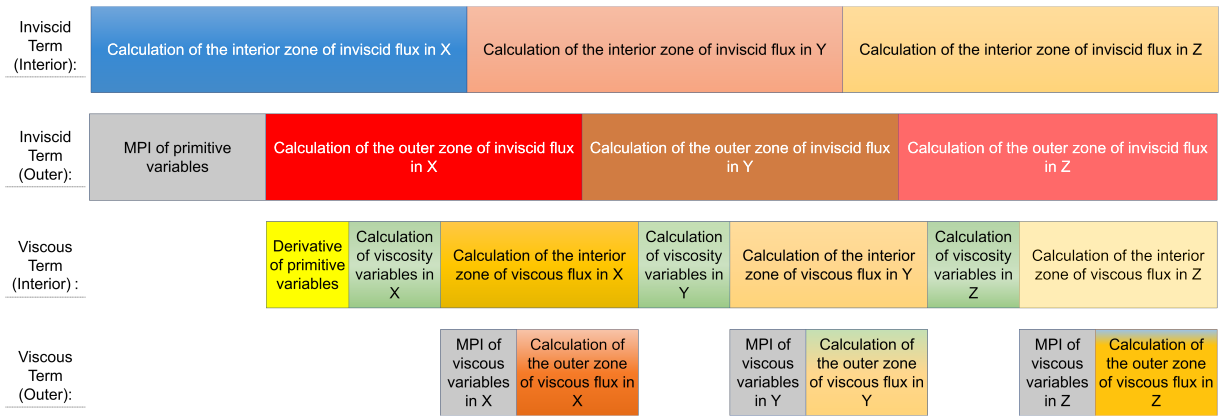

Figure 6: Program timeline design of OpenCFD-SCU.

Overlap of computing and communication has a significant effect in improving program scalability. On a CPU, MPI communication and computing overlap are usually accomplished by multithread or nonblocking instructions of MPI. In the heterogeneous CPU–GPU architecture, because the respective instructions executed on the CPU and GPU are asynchronous, those executed on the CPU will not interfere with the operations on the GPU. As long as the amount of computation undertaken by the GPU is large enough, regardless of whether the MPI communication on the CPU is blocked, the CPU will complete the MPI data exchange before the GPU acquires the data. At this time, the overlap of computing communication of the program only needs to consider the overlap of computing on the GPU and communication between CPU and GPU, which is usually realized by CUDA-Stream technology, which is dedicated to task-level concurrency on GPUs. In our design, there are four streams on the GPU, as shown in Fig. 6. After the start of the flux calculation, the calculations of the inner and outer zones of the inviscid term are placed on two streams. Only the outer zone calculation needs to obtain the adjacent process data, and this can be executed concurrently. After data exchange of the original variables in the outer region of the inviscid term has been completed, the calculations of the interior and outer regions of the viscous term are started on the other two streams. Data exchange of viscous variables between processes is needed before the viscous outer term is calculated, and the data communication also overlaps with the calculations of other streams. For the actual profiling snapshot, see Section 4.

### 3.4. IO optimization

In large-scale parallel computing, input and output (IO) may cause a bottleneck owing to the large amount of data being processed. Many high-performance computing applications let each process read and write a data file separately, although this is optional. However, if the number of MPI processes is adjusted, this can lead to trouble when computation is continued. MPI provides parallel IO interface functions that allow all MPI processes to read and write to a file at the same time. We use this function to design the IO part of OpenCFD-SCU. When the data are read, we use the Application Program Interface (API) function *MPI_File_read_at()*. The same line gives the data of one line in the file through collective communication and then chooses which to retain according to index information. When the data are written out, each process transfers data to the first process in this line, which is summed up, and the API function *MPI_File_write_at()* is called to write out the data, reducing the number of *MPI_File_write_at()* function calls. Moreover, this approach avoids the need for waiting in the same line when writing the file. This optimization of IO allows the flow field and grid of tens of billions of grid problems to be read in just ten minutes.

## 4. GPU acceleration and optimization

This section introduces the hardware environment, implementation, and optimization techniques of OpenCFD-SCU using GPU acceleration. The most significant difference between GPU programming and CPU programming is that with the former a fine-grained multithreaded programming method is adopted, requiring programmers to be more familiar with the hardware structure to achieve better performance. OpenCFD-SCU makes each GPU thread responsible for calculating a grid point by traversing multiple blocks to complete the calculation of all grid points. The optimization part mainly focuses on GPU data access. With the use of shared memory, warp shuffle, and other technologies, the performance of the kernel function is more than doubled before and after optimization, and the program has also been optimized with respect to CPU and GPU data transmission and the calculation of the kernel function.

### 4.1. Hardware

OpenCFD-SCU is heterogeneous software with cross-platform capability. It is written in CUDA-C. When writing the program, we mainly debugged it on the NVIDIA platform. However, most of the cases and tests in this paper are implemented on a supercomputer with AMD Radeon Instinct MI60 as the GPU accelerator. CUDA code cannot be compiled and run directly on an AMD platform. The programming language supported by AMD GPUs is HIP, but AMD also provides a transcoding tool called HIPIFY that can translate CUDA code to HIP [22]. When writing programs, we have avoided using API functions, such as *CUDAMemcpyToSymbol*, for which there is not an exact correspondence between CUDA and HIP. Therefore, this set of programs can not only be run directly on the NVIDIA platform, but also be compiled and run using the HIPCC compiler in the AMD environment after transcoding with HIPIFY tools. It should be noted that HIPIFY does not support CUDA-Fortran transcoding, which is why we initially rewrote the Fortran version of the CPU program with CUDA-C.

The Radeon Instinct M160 GPU die size is 331 mm$^2$ and consists of $13.2 \times 10^9$ transistors with 6.144 TFLOPS double-precision floating-point power. The parameters relevant to programming and optimization of this GPU can be found in Tables 3 and 4, which show the hardware structure parameters and memory-related parameters, respectively. Further details of the parameters of this GPU can be found in [23].

Although the architecture of an AMD GPU is not the same as that of an NVIDIA GPU, and there are many other differences between hardware products from different manufacturers or different generations, GPUs do have much in common in their architecture. They also have much in common with regard to programming. A GPU has many computing cores, organized in the form of SIMT (single instruction, multiple threads). The most basic unit for executing tasks on a GPU is not a single thread, but rather a group of multiple threads called a warp. At any time, the threads in a warp will execute the same instruction simultaneously, although they can process different data. Multiple warps will also be packaged as a block, and the programmer mainly controls the behavior of the block. The size of a block generally depends on the resource occupancy of the kernel function, although it generally remains unchanged during a calculation. The threads in a block are mapped to the specific hardware of the GPU: this hardware structure is called the compute unit (CU) in the AMD GPU architecture and the streaming multiprocessor (SM) in the NVIDIA GPU architecture (see Fig. 7), and the threads in a block will share the shared resources in the CU or SM, such as registers, shared memory, and the bandwidth of global memory.

Table 3: Computing capability.

| Model | Threads per multiprocessor | Threads per warp | Warps per multiprocessor |
| --- | --- | --- | --- |
| Radeon Instinct M160 (Vega 20 GL) | 4096 | 64 | 64 |

Table 4: Memory-related parameters.

| Model | Memory bus type | Memory bus width | Memory size | Shared memory bus width |
| --- | --- | --- | --- | --- |
| Radeon Instinct MI60 (Vega 20 GL) | HBM2 | 4096 bits | 16 GB | 128 bits |

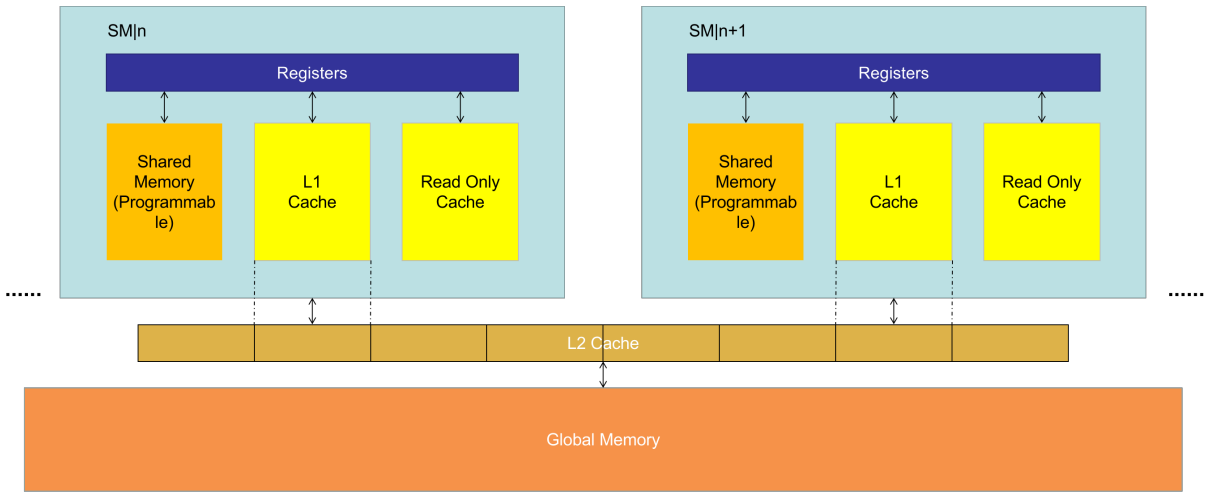

Figure 7: Memory structure of the GPU.

A GPU can start calculations of multiple blocks at the same time, mainly depending on the number of CUs or SMs on the GPU. A block will traverse all the data on the computing grid, with the grid being the total data that the kernel function needs to deal with, while the size of the data scale that the GPU can hold depends mainly on how much global memory is available. When the warp needs data, the CU or SM obtains global memory data through a cache. The L1 and L2 caches are resources within the CU or SM. In a GPU with the AMD Vega 20 architecture, the cache line widths of the L1 and L2 caches are 256 bits, the type of global memory is HBM2, there are eight dies per stack, and the total bit width of the global memory is 4096 bits. For each stack, 512 bits of data are accessed at one time.

### 4.2. Implementation

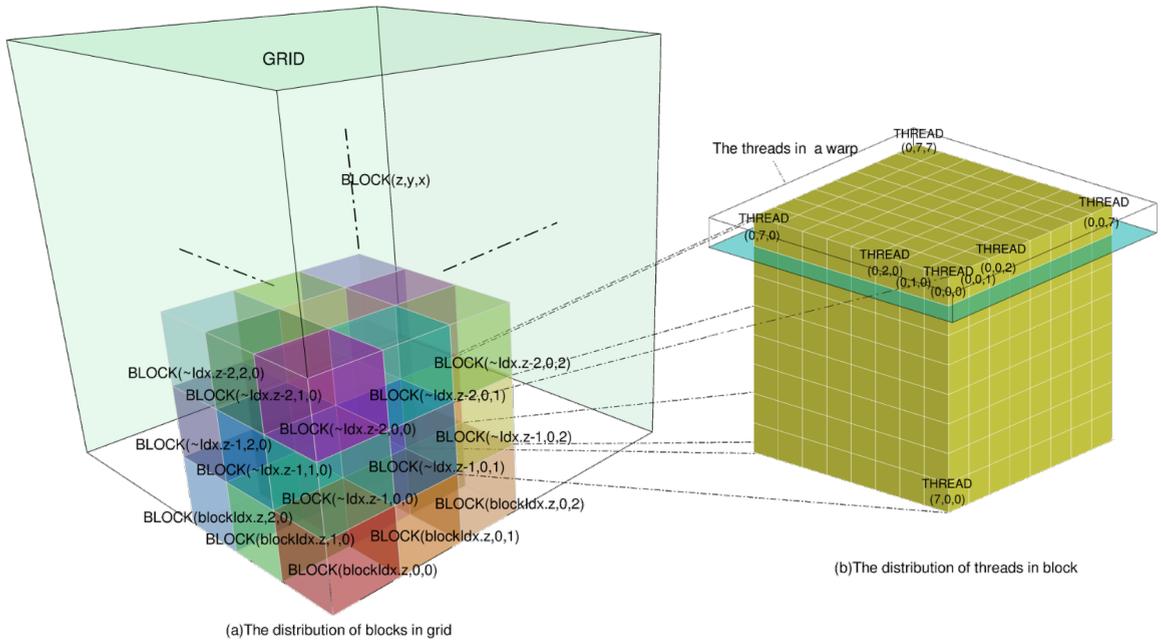

Figure 8: Three-dimensional block used by OpenCFD-SCU.

Blocks can be one-, two-, or three-dimensional. For OpenCFD-SCU, Fig. 8 shows the pattern in which the threads in a GPU are organized in a three-dimensional block. Each thread is responsible for calculating only one grid point. Each block responsible for calculating a small three-dimensional zone in the grid contains $8 \times 8 \times 8$ or $8 \times 8 \times 4$ threads in OpenCFD-SCU. The first two dimensions of the block are set to $8 \times 8$ because the warp size on an AMD M160 is 64, and this setting allows the best utilization of global memory bandwidth for a warp process with precisely one layer of data. For a GPU with 16 GB global memory, we have tested the maximum computing task with a grid size of $280^3$. The programming kernel function can obtain information about the block dimension through the embedded variables blockDim.x, blockDim.y, and blockDim.z, information about the block index through blockIdx.x, blockIdx.y, and blockIdx.z, and information about the thread index through threadIdx.x, threadIdx.y, and threadIdx.z. The program implementation process can be simplified as shown in Fig. 9.

As already mentioned, the kernel performs all floating-point operations in OpenCFD-SCU on the GPU, with the CPU being responsible only for data exchange and logic control.

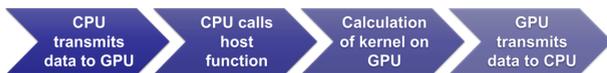

Figure 9: Flow diagram of heterogeneous program implementation.

### 4.3. Optimization

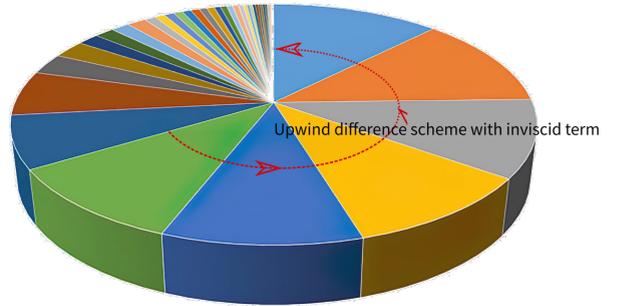

Figure 10: Hot spot analysis.

To achieve the best performance of the program, we optimize it according to the characteristics of the GPU hardware. Fig. 10 shows the proportion of the running time associated with each part of OpenCFD, which provides us with a better understanding of which parts of the program we should optimize. OpenCFD-SCU is a differential program, with the differential calculations accounting for the vast majority of the total program time. In particular, the upwind difference of the inviscid term accounts for 70% of the total program time. Hence, program optimization focuses on the differencing of the inviscid term. Differential computing is a typical stencil problem.

Like other stencil problems, the most important bottleneck is represented by the memory bound. The overall aim of memory access optimization is to ensure coalesced access to global memory and to reduce as much as possible the number of times the global memory is accessed.

### 4.3.1. Coalesced access to global memory

The first thing to consider with regard to memory access optimization is coalescing global memory access. The high bandwidth of GPU memory is realized by a high bit width; that is, a memory transaction can obtain a line of data with contiguous addresses. In the GPU that we use, this bit width is 128 bytes. If the access is coalesced and aligned, in blockIdx.x is 8, it takes only one memory transaction to obtain the data (16 double-precision data) needed by one line of threads in the block for differential calculation, as shown in Fig. 11.

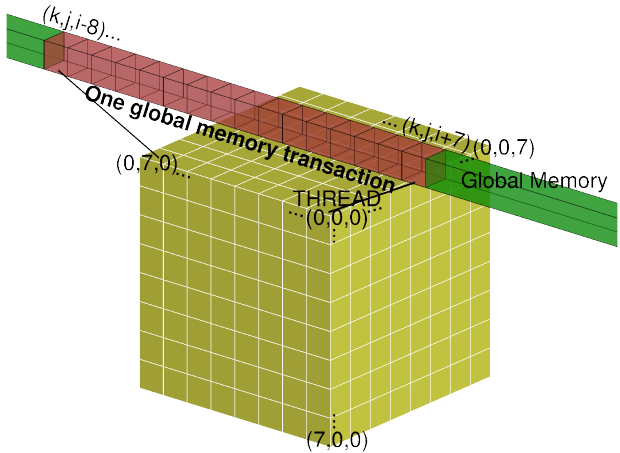

Figure 11: Intra-block thread acquires data.

Because the memory block is three-dimensional, the cudaMemalloc3d API can be used when applying for memory, which can fill in some data, thus ensuring that the starting position of each row of memory is the beginning of a memory transaction. The way to implement coalesced access is to have index A contiguous threads in a block look for data from global memory in x order. In OpenCFD-SCU, the data are divided into an interior zone and an outer zone to implement the calculation–communication overlap. The width of the outer area is set to 8 bytes, and this setting can ensure greater utilization of global memory bandwidth when accessing data.

Coalesced access is essential for GPU bandwidth efficiency. However, in a GPU, coalesced access cannot achieve 100% efficiency of GPU global memory bandwidth, and the problem of data access alignment should also be considered; that is, the starting data required by each warp should be precisely the beginning of a memory transaction access in global memory. Because the width of the lap of the difference stencil is less than the width of the global memory transaction access data, it is difficult to align the access of each warp fully. However, the L2 cache can play some role in this task, and the duplicate data obtained between different blocks can be placed in the L2 cache.

### 4.3.2. Warp shuffle and redundant computing

When OpenCFD-SCU calculates the upwind difference, the difference scheme is calculated and reorganized in form conservation form: adjacent threads calculate $\hat{f}_{j-1/2}$ and $\hat{f}_{j+1/2}$, respectively, and the derivative $\partial f/\partial x = (\hat{f}_{j+1/2} - \hat{f}_{j-1/2})/\Delta x$. This can reduce the total number of differential computations, but it does require data exchange between threads. In the program, data exchange in a block is carried out using warp-shuffle, a data exchange instruction that can rapidly perform data exchange between neighboring threads or specified threads in a warp. In the GPU of the supercomputer that we use, the warp size is 64, and we set blockdim to 8:8:4, which avoids data exchange between warps within a block. Each block can calculate the flux $f_j$ of 7:8:4 points. Warp shuffle of double-precision data is not supported by the GPU of the supercomputer that we use. Here, we convert 8-byte double-precision data into two 4-byte data through forced type conversion and perform two warp-shuffle operations. After data exchange, two 4-byte data are reorganized into double-precision data through forced type conversion.

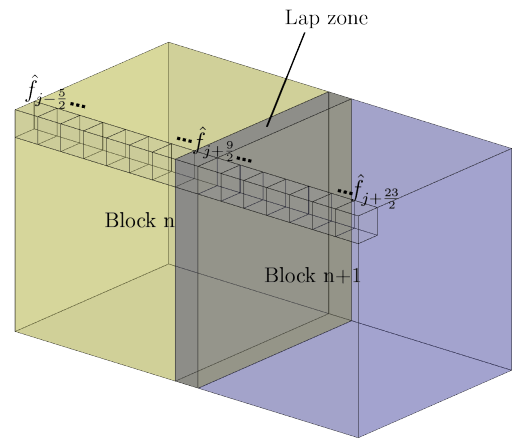

Figure 12: Redundancy zone between blocks.

$\hat{f}_{j-1/2}$ and $\hat{f}_{j+1/2}$ cannot be exchanged between blocks through warp shuffle. The usual way of dealing with this to write back to global memory, synchronize the kernel, and then access it from global memory, which involves considerable overhead. Our choice instead is to perform some redundant calculations. As shown in Fig. 12, block *n* calculates data from $\hat{f}_{j-5/2}$ to $\hat{f}_{j+9/2}$, and block (*n*+1) calculates data from $\hat{f}_{j+9/2}$ to $\hat{f}_{j+23/2}$. Although all the blocks calculate $\hat{f}_{j+9/2}$, data exchange between blocks is avoided. Furthermore, because a memory transaction accesses the data, there is no additional overhead incurred by accessing the global memory.

### 4.3.3. Thread rearrangement using shared memory

As described in Sections 4.3.1 and 4.3.2, threads in a block need to coalesce their access to global memory, and different threads in a warp must exchange $\hat{f}_{j-1/2}$ and $\hat{f}_{j+1/2}$ through warp shuffle. However, for the *y* and *z* directions, when their access is coalesced, the data accessed by adjacent lane ID threads in a block are not adjacent to the difference stencil, and, for the use of the warp shuffle, the data in a warp need to be transposed. A shared memory implementation is used here, where index threadIdx.y adjacent threads write data to the adjacent shared memory address. The index threadIdx.x adjacent thread fetches data from the shared memory. After the data have been transposed within the warp, the thread ID needs to be rearranged accordingly, as shown in Fig. 13.

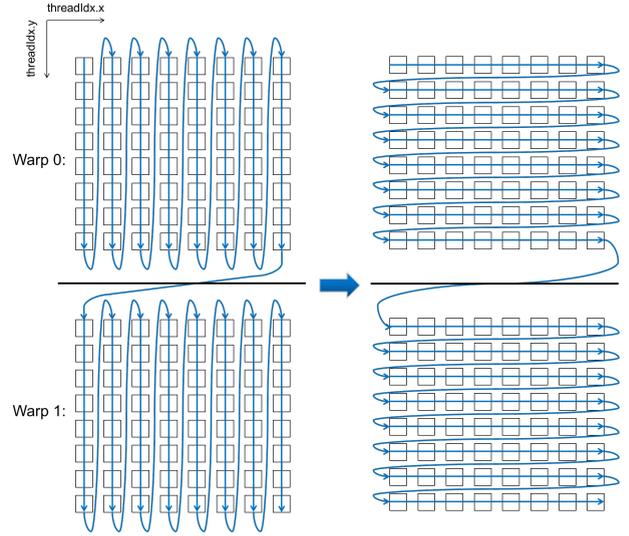

Figure 13: Data acquisition mode and thread rearrangement in a warp.

Because the warp size of the GPU on the supercomputer that we use is 64, and the transposed and rearranged matrix within each warp happens to be an 8×8 matrix, synchronization is not required when using shared memory, thus avoiding the overhead incurred by thread synchronization. However, on GPUs with other architectures, the warp size is not 64, and thread synchronization is required before accessing data from shared memory. The specific code is as follows:

```
unsigned int x = coords->x = blockDim.x * blockIdx.x + threadIdx.x;
unsigned int y = coords->y = (blockDim.y-1) * blockIdx.y + threadIdx.y;
unsigned int z = coords->z = blockDim.z * blockIdx.z + threadIdx.z;

unsigned int ID1 = 128*threadIdx.z + 16*threadIdx.x + threadIdx.y;
unsigned int ID2 = 128*threadIdx.z + 16*threadIdx.y + threadIdx.x;

shared_memory[ID1] = get_data(f, x, y-LAP+1, z, num, offset);

if(x < (job.end.x-job.start.x) && y < (job.end.y-job.start.y-1) &&
z < (job.end.z-job.start.z))
    shared_memory[ID1+8] = get_data(f, x, y+LAP+1, z, num, offset);

for(int i = ka1; i <= kb1; i++){
    stencil[i-ka1] = shared_memory[ID2+i+3];
}

x = coords->x = blockDim.x * blockIdx.x + threadIdx.y;
y = coords->y = (blockDim.y-1) * blockIdx.y + threadIdx.x;
```

Fig. 14 shows the running timeline of the program before and after coalesced access and thread rearrangement in all three directions. The grid size is $280^3$. It can be seen that after optimization, the running times of the inviscid difference kernel function WENO-SYMBO in the *y* and *z* directions are the same as that in the *x* direction, and the total running time of the program is 23.4% faster than that before optimization. In comparison, the calculation speed of the single kernel function in the *z* direction is 60% faster.

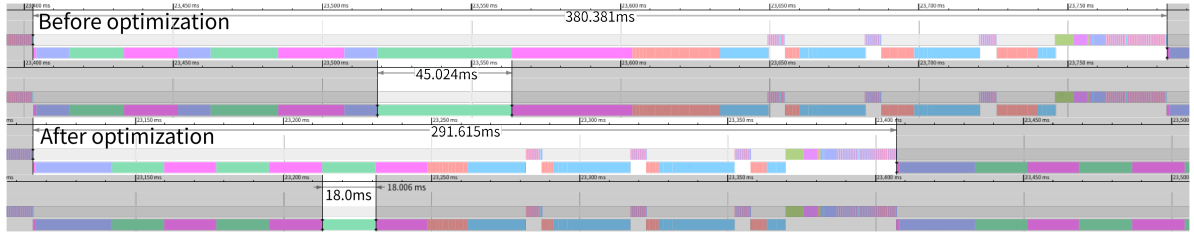

Figure 14: Program timeline before and after optimization.

### 4.3.4. Atomic operations to avoid memory access errors

When different warps on the GPU access the same memory address, memory access conflicts will occur. When getting data, conflicts will not cause errors. However, conflicts may lead to errors in the final results when different threads write data to the same address, such as when adding some values to the same memory address. Atomic operations can solve these problems. OpenCFD calculates the inviscid term and the viscous term on different streams. After these calculations, the contributed flux values are added to the same memory address. To avoid errors in the final calculation, the atomic function *__atomicAdd()* is used for accumulation in the program.

### 4.3.5. Data packaging and pinned memory for CPU–GPU data exchange

Because all the calculations in OpenCFD-SCU are carried out on the GPU, data MPI-exchange between processes needs first to transfer data from the GPU to the CPU, then transfer the exchanged data from the CPU to the GPU after the exchange has been completed. The data packaging is the first aspect of CPU–GPU data transmission that requires attention, because, when the MPI communicates, the data that need to be transmitted are in only a few thin outer layers, and their addresses in the memory are discontinuous. Tests have shown that the efficiency of calling CUDA is exceptionally low for discontinuous data, and so applying a one-dimensional continuous buffer memory on the GPU is necessary before data transmission. The data are packaged into the cache area on the GPU and then transferred. After completion of the MPI exchange, the packaged data obtained from the adjacent process are unpacked.

In addition, the application of pinned memory on the CPU is necessary to open up multiple streams to overlap computing and communications. CUDA provides a runtime API called *cudaHostAlloc()* to apply pinned memory on the CPU.

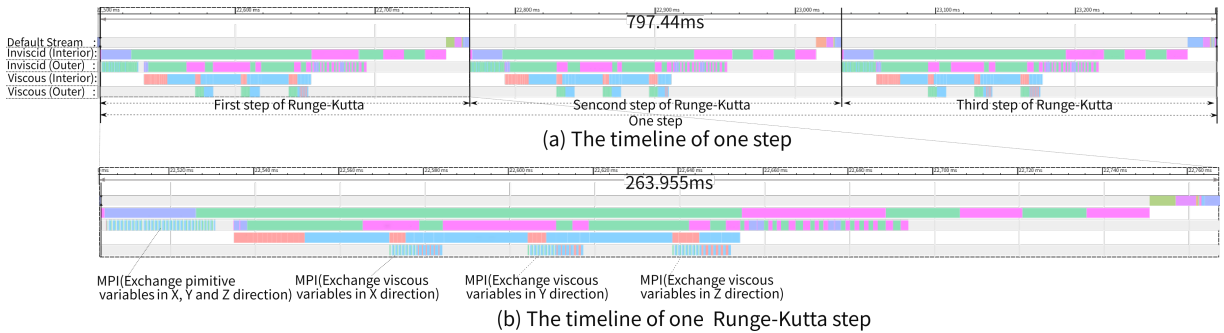

Figure 15: Program timeline in the case of multiple pipelining: (a) one step; (b) one Runge–Kutta step.

### 4.3.6. Multiple streams

We have described the design of the stream in Section 3; its primary purpose is to ensure computational communication overlap. However, it can also improve the running speed of single processes in the program. An actual profiling snapshot is shown in Fig. 15. The grid size is $280^3$. Profiling found that the actual streams are roughly the same as in the original design, and calculations on other streams can lap MPI communication. However, when multiple streams are computing tasks simultaneously, the kernel function on each stream will slow down, which can be attributed to the excessive load on the GPU. Even so, multiple streams still improve the calculation speed of the program. The running speed of the program for one Runge–Kutta step has been increased from 291 ms to 265 ms.

### 4.3.7. Other optimization techniques

In addition to the above optimization methods, the program is also optimized in other regards.

At the algorithm level, to reduce the number of global memory transactions, the Steger–Warming flux vector splitting calculation in three directions of the interior zone is written into one kernel function. Although more global memory is then needed, the access to the original variables from the global memory is reduced twofold. In the calculation of the viscous term, similar operations are carried out. The kernel functions declared by the __host__ are used for multiple packaging __device__ declared kernel functions. The intermediate variables are put into the register to avoid multiple access to global memory.

GPUs are not good at logical operations, and in program implementation, it is desirable to put judgments outside the kernel function, thereby avoiding branching within the warp. However, in some cases, a branch can reduce access to global memory, and when calculating the boundary scheme, this scheme is integrated into the difference scheme. We add a judgment to determine whether a point calculated by a thread is located at the boundary. We can then choose to use a difference or boundary scheme, which reduces the overhead.

SMs have limited numbers of registers available. When implementing kernel functions, it is desirable to reduce the number of registers used per thread and to reuse explicitly declared registers as much as possible. Although the compiler can sometimes optimize registers, it is still possible to reduce the use of some registers through appropriate algorithm design. To improve program concurrency, registers can be further optimized at the assembly level. Because the program is memory-bound, only some simple optimizations have been performed here.

In addition, the cost of calculating floating-point division in a GPU is much higher than that of multiplication. We reduce the number of division operations as much as possible in the program implementation. When division is unavoidable, we merge the division calculation. For example, the division operation of the seventh-order WENO kernel function is put on the outside. Only one division need be performed at the end of the function, although many multiplication operations are added. The running speed of the kernel function has also been made faster, and the seventh-order WENO kernel function can achieve 10% acceleration with a $280^3$ grid.

## 5. Performance and parallel scalability testing

This section mainly concerns the correctness of the program and tests of its performance and scaling. We first test correctness. Under the condition that the algorithm is the same and there is no random disturbance, it is found that the flow field file output of the CPU program is consistent with that of the GPU program. We then use the CPU and GPU programs to simulate the same case and test the acceleration, finding that the GPU program is 200–500 times faster than the CPU program. We also test both strong and weak scaling of the program: strong scaling maintains 60% parallel efficiency when the number of processes is increased by 16 times, and weak scaling maintains 98.7% parallel efficiency on 24 576 GPUs, with the grid size reaching 500 billion. At this scale, we evaluate the program performance and find that it reaches 17.1 PFLOPS, accounting for 11.4% of the GPU performance of the corresponding node.

### 5.1. Correctness test

Correctness is a prerequisite for a program to be of any practical use. OpenCFD-SCU is transplanted to a GPU from well-established CPU software called OpenCFD-SC, and we first compare it with the latter. OpenCFD-SC, written by Li, is a widely used program based on the finite difference method and has been verified in many applications to DNS of turbulence [24–27].

In the correctness test, the CPU and GPU programs calculate the same flat-plate case, using the same scheme. The grids for this calculation are set to $250 \times 240 \times 90$, i.e., $5.4 \times 10^6$, and no random disturbance is imposed. The two programs are each run for 1000 steps, and the flow field is saved. A calculation with 1000 steps is enough for the flow field data output by the programs to be very different from the initial field, and the accuracy of flow field preservation is f16.9. The DIFF tool under Linux is then used to compare the two output flow field files. It is find that there are only 56 different lines in the 5.4 million line file, with the only difference lying in positive and negative signs before the lines, which proves that the calculation results of the GPU and CPU programs are consistent.

We also use the software to simulate a classic DNS case, namely a Mach 2.9 compression ramp [7]. A more extensive and detailed calculation will be performed in Section 6.1, where a more detailed description of the case can be found. When using this case to test the correctness of the program, we adopt the same approach to calculating the inviscid term as in [7], using the WENO-SYMBO scheme, and we use an eighth-order central difference scheme for the viscous term. This test case uses $16\,100 \times 240 \times 200$ grids. However, we differ from [7] in that we develop turbulence by the addition of blowing and suction disturbances on a flat plate [5], rather than by recycling.

Fig. 16 shows the turbulence development on an $X$–$Y$ section, as represented by the instantaneous temperature, Fig. 17 shows the mean velocity profiles, and Fig. 18 compares the wall pressure with experimental results [28]. There is good agreement between the results from OpenCFD-SCU, those from the DNS in [7], and experiment, thus verifying the correctness of the program.

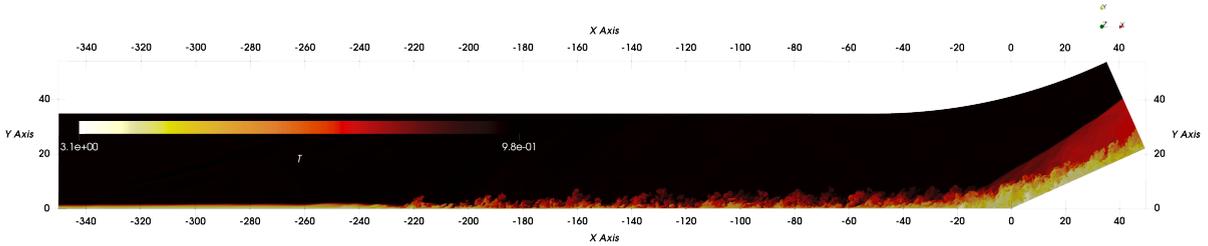

Figure 16: Instantaneous temperature, nondimensionalized by the reference temperature.

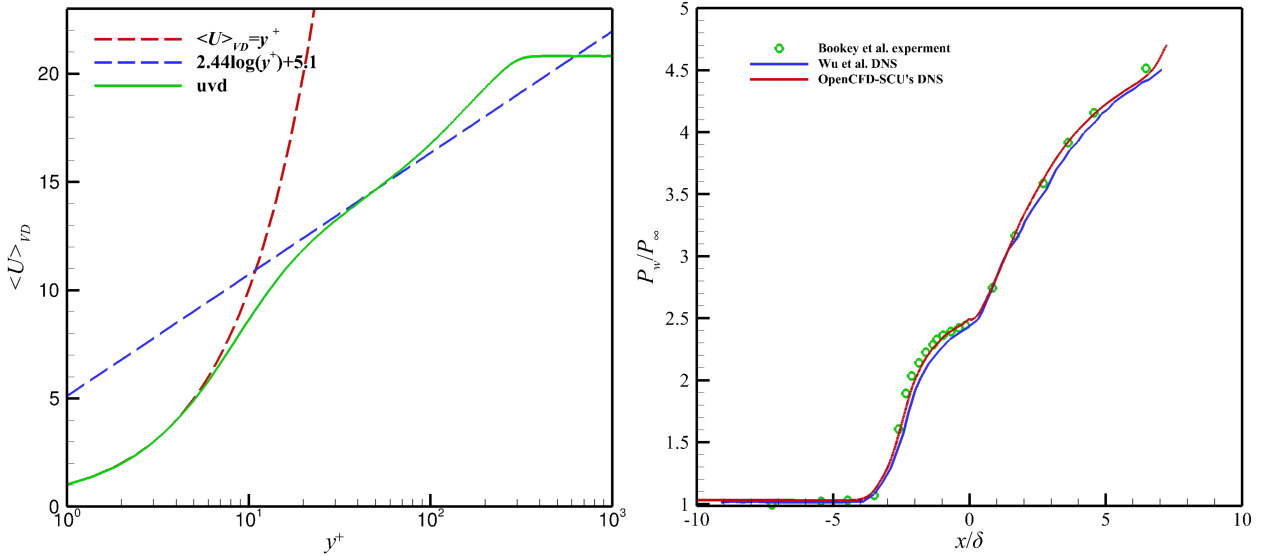

Figure 17: Van Driest transformed mean velocity profiles.

Figure 18: Mean wall-pressure distribution from DNS and experimental data, with 3% error bars.

### 5.2. Acceleration compared with CPU program

Table 5: Comparison of times taken by OpenCFD-SC and OpenCFD-SCU when simulating the same problem.

| Program | Number of MPI processes | Time per step (s) |
|---|---|---|
| OpenCFD-SC | 1280 | 10.66 |
|  | 9 120 | 1.93 |
|  | 36 000 | 0.77 |
|  | 72 576 | 0.54 |
|  | 131 712 | 0.39 |
| OpenCFD-SCU | 128 | 0.5174 |
|  | 256 | 0.2708 |
|  | 512 | 0.1387 |
|  | 1 024 | 0.0856 |
|  | 2 048 | 0.0627 |

We use OpenCFD-SC and OpenCFD-SCU for DNS of the same case to test the acceleration provided by the latter program. The case considered is a lifting body, the model geometry is from the Hypersonic Transition Research Vehicle designed by the China Aerodynamics Research and Development Center [29–31]. The hardware parameters of the GPU used by OpenCFD-SCU have been described in Section 4.1. OpenCFD-SC uses a Haiguang CPU with base frequency 2.4 GHz. Table 5 shows the results of the test in terms of time per step for each program to run the average time of 100 steps. It can be seen that the running speed of OpenCFD-SCU with just 512 GPUs exceeds that of OpenCFD-SC with 130 000 CPUs. For the same number of MPI processes, there is a more than 200-fold acceleration. Fig. 19 provides a more illustration of the gap in performance between the two programs. OpenCFD-SCU has improved our numerical simulation ability by two orders of magnitude.

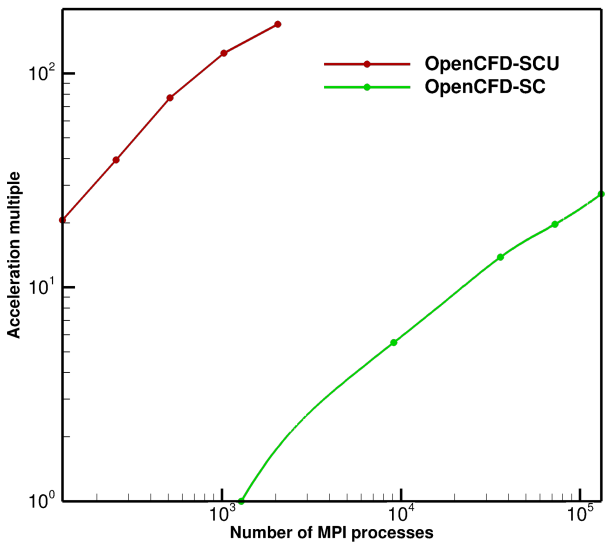

Figure 19: Comparison of calculation speed between OpenCFD-SC and OpenCFD-SCU when simulating the same problem. The base speed is selected as the speed of the minimum MPI process that OpenCFD-SC can calculate for this problem.

### 5.3. Weak scaling

We also test the weak scalability of OpenCFD-SCU software, fixing the size of the problem on each processor and observing how the solution time of the program varies with the number of processors. Each process executes a simulation task with $280^3$ grids, and the time per step is the time average of 100 steps of the program. The number of processes is varied from 1 to 24 576. It is found that the program's computing time per step remains roughly the same. When the program runs for a single process, each step of the program takes 0.791 s, whereas for the 24 576 processes the run time is 0.78 s. Fig. 20 shows the parallel efficiency of weak scalability. The parallel efficiency for the 24 576 processes reaches 98.7%. At the maximum number of processes, the program has simulated tasks with more than 500 billion grids.

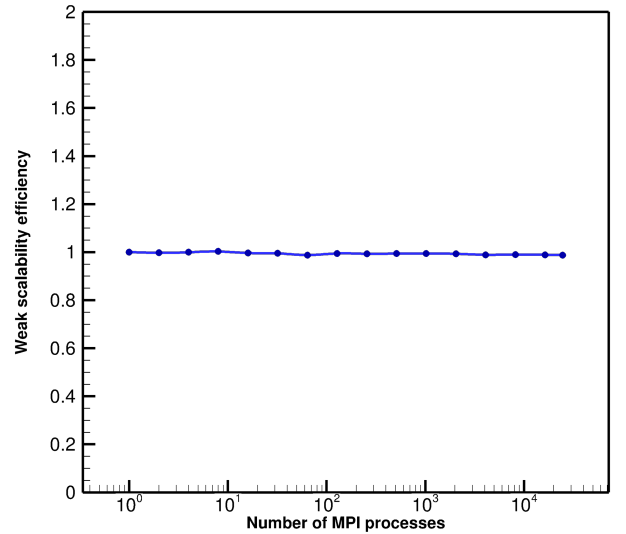

Figure 20: Weak scaling parallel efficiency.

### 5.4. Double-precision floating-point operation performance

On the basis of the above test results, we estimate the double-precision floating-point computing power of the program. Table 6 shows the number of double-precision floating-point operations of the manually counted primary functions of the program, with its double-precision floating-point performance calculated accordingly. Manual counts of the number of times have been performed only for the upper programming language visible double-precision floating-point numbers of addition, subtraction, multiplication and division operations. Addressing and other calculations have not been not counted. A floating-point operation is counted for some library functions, such as SQRT. However, GPUs do not have hardware that specifically handles this kind of library function, which must be realized instead by software. Thus, more than one floating-point operation is required for each such function, and so the number of floating-point operations shown in the table is only the base level of the program's actual number of floating-point operations. We use the software counter ROCPROF to count VALU INSTS (a parameter used to measure the number of calls to the GPU arithmetic unit) of the WENO-SYMBO kernel function on the GPU. The result is 388, which is much more than the 262 floating-point operations we counted manually, but we still use the results of the manual count to estimate the program's performance. According to our estimate, OpenCFD-SCU can reach a double-precision computing power of 705.4 GFLOPS in a single process.

Table 6: Number of double-precision floating-point operations.

| Kernel function | Kernel's floating-point operations | Calls of kernel | Grids calculated (including redundancy) | Total floating-point operations of kernel |
|---|---|---|---|---|
| WENO-SYMBO | 216 | 90 | 25 088 000 | $4.877 \times 10^{11}$ |
| Steger–Warming splitting | 138 | 9 | 22 579 200 | $2.80 \times 10^{10}$ |
| Sixth-order upwind | 9 | 72 | 21 952 000 | $1.42 \times 10^{10}$ |
| Viscous term | 128 | 9 | 22 579 200 | $2.60 \times 10^{10}$ |
| Runge–Kutta (1st + 2nd + 3rd) | 62 | 1 | 21 952 000 | $1.36 \times 10^{9}$ |
| | | Total floating point operations | | $5.5726 \times 10^{12}$ |
| | | Time | | 792 ms |
| | | Performance | | 705.4 GFLOPS |

Fig. 21 shows the performance for multiple processes. At 24 567 processes, OpenCFD-SCU can achieve 17.1p double-precision performance, accounting for 11.4% of the corresponding node supercomputing GPU performance. It can be seen that the performance of the program increases linearly with increasing number of processes.

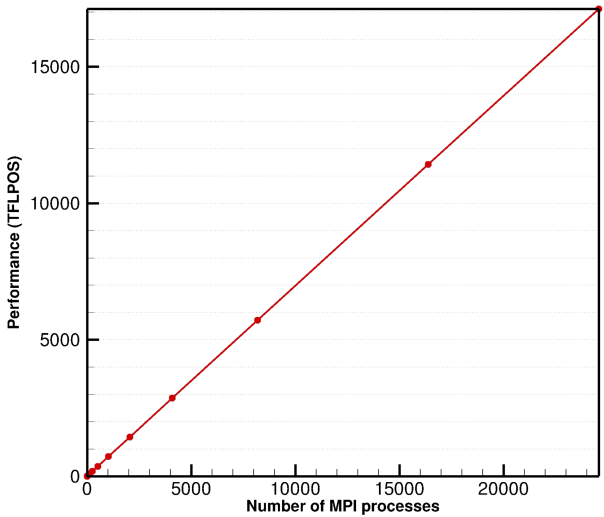

Figure 21: Variation of double-precision floating-point operation performance of OpenCFD-SCU with number of processes

### 5.5. Cross-platform testing

OpenCFD-SCU can run on both AMD and NVIDIA GPUs. All the tests described above have been carried out on the AMD platform, although we have also tested several GPUs from the latest NVIDIA platform. However, testing can only be done on a single GPU at a given time, owing to resource constraints. Table 7 shows the results of these tests. The different GPUs are used to run the same test case, with the grids set as $200 \times 200 \times 200$, and the average time taken for the program to run 100 steps is recorded. The performance of OpenCFD-SCU on the AMD GPU is much higher than those on the NVIDIA GPUs. Compared with the NVIDIA Tesla A100, the AMD Radeon M160 is nearly ten times faster, and when we compare the program output files, we find that calculation results are exactly the same. This significant gap in program performance may be due to our targeted optimization of the AMD GPU. Thus, it can be seen that thorough understanding and optimization of GPU hardware is essential to enable the program to realize its performance capabilities to the full.

Table 7: Running times of OpenCFD-SCU on different GPUs.

| GPU | Time per step (s) |
| --- | --- |
| NVIDIA Tesla V100 | 4.702 |
| NVIDIA RTX3090 | 4.038 |
| NVIDIA Tesla A100 | 3.036 |
| AMD Radeon M160 | 0.315 |

## 6. DNS of wall turbulence

With the OpenCFD-SCU program having been optimized and verified, its dramatically improved operation speed enables its application to some challenging problems. This section considers a problems that a 31.2 billion grid wide-span Mach 2.9 compression ramp with 24° ramp angle.

### 6.1. Mach 2.9 compression ramp flows

A supersonic compression ramp flow at Mach 2.9 with a wide spanwise computational domain is simulated (this is the case that was used earlier in the test in Section 5.1). The convective terms are discretized by the optimized WENO scheme proposed by Martín et al. [7], and viscous terms are discretized by an eighth-order central difference scheme. Nonreflection boundary conditions are imposed at the streamwise direction exit and on the upper boundary of the wall-normal direction before it enters the corner, and an incoming flow boundary condition is imposed after the corner area. Laminar-to-turbulent transition is triggered by the wall-normal blowing and suction disturbances proposed by Pirozzoli et al. [32]. These disturbances are imposed on the flat plate region ranging from $x = -350$ mm to $-200$ mm. An adaptive filtering technique is applied to maintain computational stability during the process of establishing the flow field [33] and is then disabled after the flow field has been established. To eliminate the influence of this filtering, further time-steps are needed after it has been disabled.

The grid setting is $16\,250 \times 240 \times 8000$ in the streamwise, wall-normal, and spanwise directions, respectively. A total of 31.2 billion grid points are used, more than 6000 times the number used by Martín et al. [7]. The mesh resolution reaches $\Delta x^+ = 1.2$, $\Delta y^+ = 0.41$, $\Delta z^+ = 1.2$ in the fully developed turbulent boundary layer. The parameters of the turbulent boundary layer are listed in Table 8. The grid is set to $16\,250 \times 240 \times 8000$ (31.2 billion), and the grid resolution is $\Delta x^+ = 1.2$, $\Delta y^+ = 0.41$, $\Delta z^+ = 1.2$ at a distance of $-35$ mm from the corner, which is taken as a reference station at which the turbulence is fully developed.

Table 8: Conditions of incoming flow .

| $Ma_\infty$ | $Re_{mm}$ | $T_\infty$ | $T_w$ |
| --- | --- | --- | --- |
| 2.9 | 5581.4 | 108.1 K | 307 K |

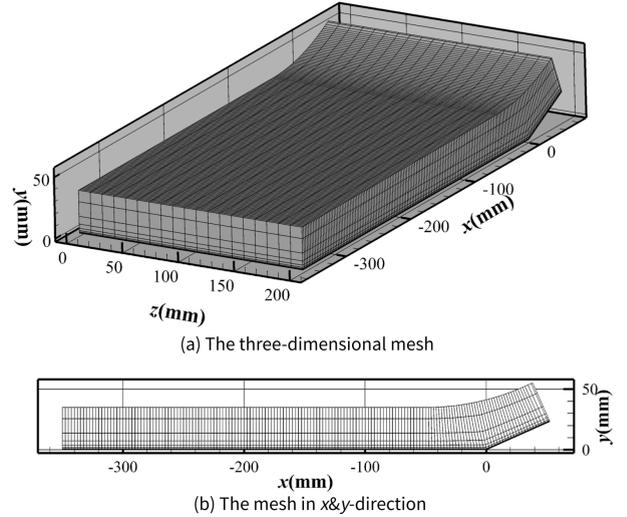

(a) The three-dimensional mesh

(b) The mesh in x&y-direction

Figure 22: Schematics of grid and computational domain: (a) three dimensions; (b) x–y section.

The distribution of the instantaneous skin friction coefficient $C_f$ is plotted in Fig. 23. Streamwise-elongated streaking structures can be observed in the upstream flat plate region, while the spanwise scales of these structures are dramatically increased in the downstream interaction region, which are closely related to Taylor–Görtler vortices [34].

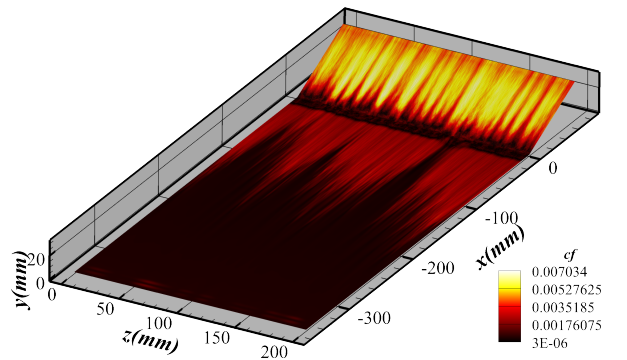

Figure 23: Distribution of skin friction coefficient $C_f$ on wall.

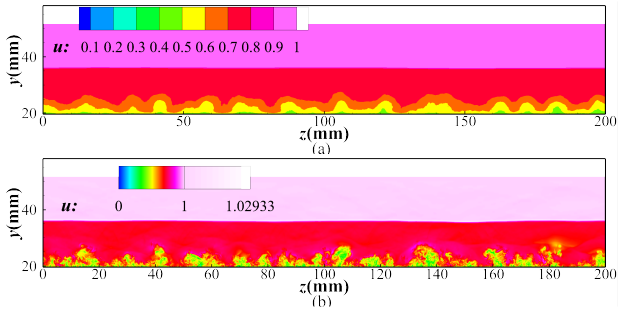

Figure 24: Velocity distribution along spanwise direction at $x = 44$ mm: (a) after time averaging; (b) instantaneous.

Fig. 24 plots the counters of the streamwise velocity in the $x$–$z$ plane, with Fig. 24(a) showing the flow field after time averaging and Fig. 24(b) an instantaneous flow field. It can be seen that a distributed Taylor–Görtler vortex structure is present after the shock wave.

The streamtraces in Fig. 25 reveal the presence of several small-scale vortices rotating in the same direction beneath a single giant Taylor–Görtler vortex.

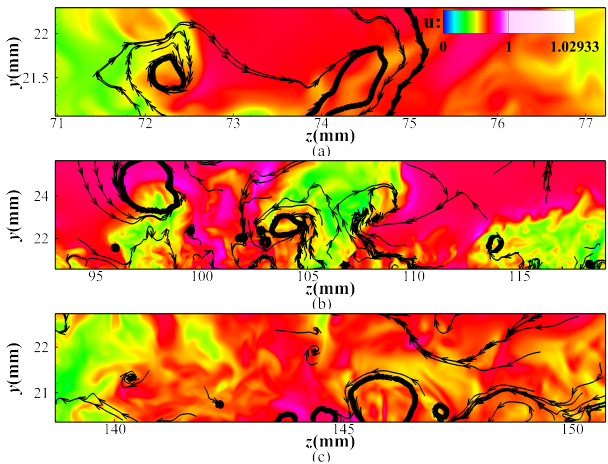

Figure 25: Instantaneous velocity and streamtraces in three different small areas.

## 7. Conclusions

This paper has introduced open source CFD software, OpenCFD-SCU, for DNS of compressible wall-bounded turbulence. This software is based on the finite difference method and is accelerated by the use of a GPU. Cross-platform deployment is possible, with the program being able to use both NVIDIA and AMD GPUs. The algorithm, heterogeneous programming, and optimization methods adopted in the program have been described here in detail. A variety of optimization techniques are applied in OpenCFD-SCU to improve its performance, including memory access coalescing, redundant computing to reduce global memory access and overlapping of computation with communication. In comparison to a CPU-based program based on the same algorithm, OpenCFD-SCU provides acceleration by a factor of more than 200 using the same MPI process. In terms of both weak and strong scaling, OpenCFD-SCU behaves well, achieving 98.7% parallel weak scalability with 24 576 GPUs. Test results indicate that OpenCFD-SCU not only accelerates the simulation speed, but also saves up to 90% of the computational cost, highlighting its suitability for large-scale DNS.

A validation and verification case of a Mach 2.9 compression ramp with mesh number up to 31.2 billion are presented in this paper. In this simulation of a turbulent boundary layer with a 24° compression ramp at Mach 2.9, the spanwise width of the computational domain was found to exceed 29 boundary layer thicknesses. In the case of a large spanwise width, a series of Taylor–Görtler vortices were observed behind the shock wave.

Actually the software includes a variety of high-precision finite difference schemes, including a seventh-order upwind scheme and seventh- and fifth-order WENO schemes. To improve computational robustness while maintaining the scheme's low dissipation and high resolution, a hybrid scheme is included using a modified Jameson shock wave recognizer. The OpenCFD-SCU source code is available at http://developer.hpccube.com/codes/danggl/opencfd-scu.git.

## ACKNOWLEDGEMENTS

This work was supported by the National Key Research and Development Program of China (2019YFA0405300) and NSFC Projects (12232018, 91852203, 12072349, 12202457).

The authors thank Sugon Advanced Computing service platform and Supercomputing Center of Chinese Academy of Sciences for providing computer time.

## Conflict of Interest

The authors have no conflicts to disclose.

## DATA AVAILABILITY

The data that support the findings of this study are available from the corresponding author upon reasonable request.